\documentstyle[prd,aps,preprint,tighten,epsf,amstex]{revtex}

\def\cancel#1#2{\ooalign{$\hfil#1\mkern1mu/\hfil$\crcr$#1#2$}}
\def\slash#1{\mathpalette\cancel{#1}}

\def\mpi2{m_\pi^2}
\def\mK2{m_K^2}

\newcommand{\bea}{\begin{eqnarray}}
\newcommand{\eea}{\end{eqnarray}}
\newcommand{\be}{\begin{equation}}
\newcommand{\ee}{\end{equation}}
\newcommand{\nn}{\nonumber}
\begin{document}
\bibliographystyle{apsrev}
\epsfclipon

\newcounter{Outline}
\setcounter{Outline}{0}

\newcounter{Intro}
\setcounter{Intro}{1}

\newcounter{Analytic}
\setcounter{Analytic}{1}

\newcounter{Scales}
\setcounter{Scales}{1}

\newcounter{PhaseTr}
\setcounter{PhaseTr}{1}

\newcounter{Karsch}
\setcounter{Karsch}{1}

\newcounter{EOS}
\setcounter{EOS}{1}

\newcounter{Conclusions}
\setcounter{Conclusions}{1}

\newcounter{Tables}
\setcounter{Tables}{1}

\newcounter{Figures}
\setcounter{Figures}{1}


\draft

\preprint{CU-TP-1126}

\title{Two-flavor QCD Thermodynamics using Anisotropic Lattices}

\author{
Ludmila~Levkova$^{a}$\footnote{\footnotesize Present address: Physics Department, Indiana University, Bloomington, IN 47405.}}
\author{
Thomas~Manke$^{a}$\footnote{\footnotesize Present address: Max Planck Insitute for Molecular Genetics, Ihnestr. 73, 14195 Berlin, Germany.}}
\author{
Robert~Mawhinney$^a$}
\address{
\vspace{0.5in}
$^a$Department of Physics, Columbia University, New York, NY, 10027}

\date{\today}
\maketitle

\begin{abstract}
Numerical simulations of full QCD on anisotropic lattices provide a 
convenient way to study QCD thermodynamics with fixed physics scales 
and reduced lattice spacing errors. We report results from calculations 
with two flavors of dynamical staggered fermions,
where all bare parameters and the
renormalized anisotropy are kept constant and the temperature is changed
in small steps by varying only the number of time slices. Including results
from zero-temperature scale setting simulations, which 
determine the Karsch coefficients, allows for the calculation of the equation
of state at finite temperatures.
\end{abstract}

\newpage


\section{Introduction}
\label{sec:intro}

\ifnum\theIntro=1
%
%

\ifnum\theOutline=1
\noindent \framebox{Begin \ outline \hspace{5.0in} }
\begin{enumerate}
\item History
\item Fixed parameters scheme
\end{enumerate}
\noindent \framebox{End \ outline \hspace{5.0in} }
\vspace{0.25in}
\fi
Lattice calculations provide a method to study the properties of the quark-gluon plasma (QGP), which is considered
to be the phase  matter existed in at the extreme range of temperatures microseconds after the big bang.  
To understand the basic character of the QGP we need to determine the equation of state (EOS) for the system in
a regime of strong gauge coupling for which a non-perturbative scheme for calculation is most adequate. Currently
in the lattice formulation of QCD two different approaches to that problem could be adopted.
One of them is the derivative (operator) method which requires the knowledge of the asymmetry coefficients, 
or Karsch coefficients \cite{k_coeff,diff_method}.
These coefficients have been evaluated only perturbatively \cite{karsch_pert,karsch_pert1} 
since their non-perturbative values are not trivial to calculate in practice \cite{isotr_res}.
The second method is the integral method \cite{int_method,int_method1}, which does not require the values
of the Karsch coefficients, but has the disadvantage 
that for a given quark mass at a single temperature a number of different simulations are required, and
in addition, there exist the problems of scaling violation. 
In our study we avoid these disadvantages by choosing the derivative method implemented for anisotropic
lattices in combination with a fixed parameter scheme described below.

The anisotropic formulation of lattice QCD has certain advantages regarding the
study of the equation of state at various finite temperatures. 
Finite temperature field theory has a 
natural asymmetry which makes the anisotropic approach useful to reduce the 
lattice spacing errors associated with the transfer matrix at less cost than 
is required for the full continuum limit \cite{anis_lat}.
Through the introduction of anisotropy on the lattice one can make the temporal lattice spacing,
 $a_t$, sufficiently small so that by varying the number of time slices, $N_t$, the temperature 
can be changed in small discrete steps. 
 
To  study the thermodynamic properties of the quark-gluon system we simulate QCD
with two flavors of dynamical staggered fermions on anisotropic 
lattice. 
The fixed parameter 
 scheme we employ to avoid scaling violations is the following.
 All the bare parameters of the simulation are kept constant and only the 
temperature is changed by varying $N_t$ (from 4 to 64). This approach separates temperature and lattice 
spacing effects and keeps the underlying physics scales (i.e. the lattice spacings in
 temporal and spatial directions, respectively $a_t$ and $a_s$) fixed.

The calculation of EOS of the quark-gluon system involves derivatives 
of the bare parameters with respect to the physical anisotropy $\xi = a_s/a_t$ and the spatial lattice spacing 
$a_s$. These are the already mentioned above 
Karsch coefficients, which can be obtained non-perturbatively as a by-product of the zero-temperature
calculations needed to choose the bare parameters. In our scheme, once calculated, the Karsch coefficients
can be used for all temperatures since they depend only on the intrinsic lattice
parameters and not on $N_t$.  This allows a straightforward determination 
of the temperature dependence of the energy and pressure, again at fixed lattice spacing. 
With two or more slightly different values for $a_t$, a high-resolution 
sampling of temperatures can be investigated.


This paper is organized as follows. In section II, we define the anisotropic staggered action and 
derive analytic expressions for the energy and pressure for the studied system. In section III we describe 
the scale-setting techniques. Section IV investigates the phase transition for the staggered 
fermions as we change the temperature in small steps. In section V we present the technique used 
to calculate the Karsch coefficients and the numerical results for them. Section VI contains the 
numerical results for the EOS for two anisotropies, $\xi= 4.0(1)$ and $\xi= 4.8(3)$. 
In Section VII we examine our data for evidence for improvement of the flavor symmetry,
when due to the anisotropy $a_t$ becomes sufficiently small. 
Our conclusions are given in section VIII.

\fi


\section{Action and General Analytic Framework}
\label{sec:analytic}

\ifnum\theAnalytic=1
%
%

\ifnum\theOutline=1
\noindent \framebox{Begin \ outline \hspace{5.0in} }
\begin{enumerate}
\item Action
\item EOS
\end{enumerate}
\noindent \framebox{End \ outline \hspace{5.0in} }
\vspace{0.25in}
\fi

In this section we define the action we use and derive the analytical form of the EOS using the derivative method.

We work with an asymmetric lattice in Euclidean space with notation for 
the spatial lattice spacing $a_s$ and temporal 
lattice spacing $a_t$. Our calculations are based on the QCD action
$S = S_G +  S_F$,  where the gauge part is the standard anisotropic Wilson action \cite{wil_action}:
\begin{eqnarray}
\label{eq:asym_gauge_action}
S_G
 &=&- \frac{\beta}{N_c} \left[ \frac{1}{\xi_o}
     \sum_{x, s>s^\prime} {\rm Re} {\rm Tr}\left[ P_{ss^\prime}(x)\right] 
     + \xi_o \sum_{x, s} {\rm Re} {\rm Tr}\left[P_{st}(x)\right] 
     \right]
\end{eqnarray}
and the fermion part is the standard staggered action \cite{stagg_action} with anisotropy 
introduced in the spirit of \cite{wil_action} :
\begin{eqnarray} 
\label{eq:aniso_lattice_quark_action}
S_F
  &=& \sum_{x} \overline{\psi}(x) \left[
      m_f + 
      \nu_t  \slash{D}_t + \frac{\nu_s}{\xi_o} \sum_s \slash{D}_s 
      \right]  \psi(x) . 
\end{eqnarray}

In the above definitions $P_{ss^\prime}$ and $P_{st}$ are the space-space and space-time
plaquette variables. The bare anisotropy parameter $\xi_o=\xi$ at the tree level. $\slash{D}_t$ and $\slash{D}_s$ 
are the temporal and spatial parts of the staggered Dirac operator, $\nu_t$ and $\nu_s$ are 
the bare speed of light parameters and $N_c=3$ is the number of colors.

After integrating out the fermion fields in the path integral explicitly, the  
fermion action effectively becomes:
\begin{eqnarray} 
\label{eq:aniso_lattice_quark_action_eff}
S_F&=& -\frac{N_f}{4}{\rm Tr}( \ln M),
\eea
where $N_f=2$ is the number of fermion flavors and $M$ is the fermion matrix which has the form:
\bea
M&=& m_fI +\nu_t\slash{D}_t+\frac{\nu_s}{\xi_o}\slash{D}_s.
\end{eqnarray}
We choose $\nu_s=1$ at the expense of rescaling the quark fields and in an actual simulation $\nu_t$ 
is tuned so that the relativistic properties of the action are restored.

To determine the energy density $\varepsilon(T)$ and pressure $p(T)$ as functions of the temperature T we use the thermodynamic identities:  
\begin{eqnarray}
\varepsilon(T)&=&-\frac{1}{V_s}\left.\frac{\partial\ln Z}{\partial (1/T)}\right|_{V_s}\\
p(T)&=&T\left.\frac{\partial\ln Z}{\partial V_s}\right|_{T},
\end{eqnarray}
where the partition function is $Z=\int\left[ d\psi d\overline{\psi} dU\right]\exp(-S)$, the volume is 
$V_s=N_s^3a_s^3$ and $1/T=N_ta_t$. $N_s$ and $N_t$ are respectively the number of spatial and temporal lattice sites.
This way we have:
\begin{eqnarray}
\label{eq:7}
\varepsilon(T)&=& -\frac{\xi}{N_s^3N_ta_s^3a_t}\langle\left. \frac{\partial S}{\partial \xi}\right|_{a_s} \rangle\\
\label{eq:8}
p(T)&=&-\frac{a_s}{3N_s^3N_ta_s^3a_t}\langle\left.\frac{\partial S}{\partial a_s}\right|_{a_t}\rangle, 
\end{eqnarray}
where the angle brackets denote averaging over the gauge ensemble.

The physical anisotropy is defined as $\xi=a_s/a_t$. It is convenient to choose $\xi$ and $a_s$ to  be the independent 
variables in this relation. This allows Eq.~(\ref{eq:7}) and Eq.~(\ref{eq:8}) 
to be written only in terms of derivatives of them via the transformations:
\begin{eqnarray}
\left.\frac{\partial}{\partial a_t}\right|_{a_s}&=&-\frac{a_s}{a_t^2}\left.\frac{\partial}{\partial \xi}\right|_{a_s}\\
\left.\frac{\partial}{\partial a_s}\right|_{a_t}&=&\left.\frac{\partial}{\partial a_s}\right|_\xi + \frac{1}{a_t}\left.\frac{\partial}{\partial \xi}\right|_{a_s}.
\end{eqnarray}
Thus the expression for the pressure $p(T)$ becomes: 
\begin{eqnarray}
\label{eq:11}
p(T) &=&-\frac{a_s}{3N_s^3N_ta_s^3a_t}\left[\langle\left.\frac{\partial S}{\partial a_s}\right|_{\xi}\rangle +
\frac{1}{a_t} \langle\left.\frac{\partial S}{\partial \xi}\right|_{a_s}\rangle\right]\\
&=&-\frac{\xi}{3N_s^3N_ta_s^3a_t}\langle\left.\frac{\partial S}{\partial \xi}\right|_{a_s}\rangle 
- \frac{a_s}{3N_s^3N_ta_s^3a_t}\langle\left.\frac{\partial S}{\partial a_s}\right|_{\xi}\rangle \nonumber\\
&=&\frac{\varepsilon(T)}{3}-  \frac{a_s}{3N_s^3N_ta_s^3a_t}\langle\left.\frac{\partial S}{\partial a_s}\right|_{\xi}\rangle.\nn
\end{eqnarray}
To simplify the analytic expressions, in the explicit form of the derivatives of the action $S$ 
we use the following normalization notations:
\begin{eqnarray}
\langle{\rm Re} {\rm Tr}\left[ P_{ss^\prime}\right]\rangle&=&\frac{\langle\sum_{x, s>s^\prime} {\rm Re} {\rm Tr}\left[ P_{ss^\prime}(x)\right]\rangle}{3N_s^3N_tN_c}\\
\langle{\rm Re} {\rm Tr}\left[ P_{st}\right]\rangle&=&\frac{\langle\sum_{x, s} {\rm Re} {\rm Tr}\left[ P_{st}(x)\right]\rangle}{3N_s^3N_tN_c}\\
\langle\overline{\psi}\psi\rangle&=&\frac{\langle{\rm Tr}\left[M^{-1}\right]\rangle}{N_cN_s^3N_t}\\
\langle\overline{\psi}\slash{D}_t\psi\rangle&=&\frac{\langle {\rm Tr}\left[\slash{D}_tM^{-1}\right]\rangle}{N_cN_s^3N_t}\\
\langle\overline{\psi}\slash{D}_s\psi\rangle&=&\frac{\langle {\rm Tr}\left[\slash{D}_sM^{-1}\right]\rangle}{N_cN_s^3N_t}.
\end{eqnarray}

The equations for the energy density and pressure at a given temperature, Eq.~(\ref{eq:7}) and Eq.~(\ref{eq:11}), 
are not corrected for the zero
temperature divergent contribution, which simply should be subtracted. This subtraction is trivial and from here on 
the formulae will assume that $\varepsilon(T)$ and $p(T)$ have that correction. 
Dividing Eq.~(\ref{eq:7}) and Eq.~(\ref{eq:11}) by $T^4$ (i.e. multiplying them by $N_t^4a_t^4$) 
and using the notations from above we obtain the following final formulae:
\begin{eqnarray}
\label{eq:23}
\frac{\varepsilon(T)}{T^4}&=&\frac{3N_t^4}{\xi^2}\left[\left(\frac{1}{\xi_o}\left.\frac{\partial \beta}{\partial \xi}\right|_{a_s}+\beta\left.\frac{\partial \xi_o^{-1}}
{\partial \xi}\right|_{a_s}\right)\langle {\rm Re} {\rm Tr}\left[ P_{ss^\prime}\right]\rangle\right. \nonumber\\
&&\left.+\left(\xi_o\left.\frac{\partial \beta}{\partial \xi}\right|_{a_s}+\beta\left.\frac{\partial\xi_o}{\partial \xi}\right|_{a_s}\right)\langle
{\rm Re} {\rm Tr}\left[ P_{st}\right]\rangle \right] \nonumber\\
&&+\frac{3 N_t^4N_f}{4\xi^2}\left[\left. \frac{\partial m_f}{\partial \xi}\right|_{a_s}
\langle\overline{\psi}\psi\rangle
+\left. \frac{\partial \nu_t}{\partial \xi}\right|_{a_s}\langle\overline{\psi}\slash{D}_t\psi\rangle\right.\nn\\ 
&&+\left.\left.\frac{\partial \xi_o^{-1}}{\partial \xi}\right|_{a_s}\langle\overline{\psi}\slash{D}_s\psi\rangle\right]
\end{eqnarray} 
\begin{eqnarray}
\label{eq:24}
\frac{p(T)}{T^4}&=&\frac{\varepsilon(T)}{3T^4} + \frac{a_sN_t^4}{\xi^3}\left[\left(\frac{1}{\xi_o}\left.\frac{\partial \beta}{\partial a_s}\right|_{\xi}+\beta\left.\frac{\partial \xi_o^{-1}}
{\partial a_s}\right|_{\xi}\right)\langle  {\rm Re} {\rm Tr}\left[ P_{ss^\prime}\right]\rangle\right. \nonumber\\
&&+\left.\left(\xi_o\left.\frac{\partial \beta}{\partial a_s}\right|_{\xi}+\beta\left.\frac{\partial\xi_o}{\partial a_s}\right|_{\xi}\right)\langle
{\rm Re} {\rm Tr}\left[P_{st}\right]\rangle\right]\nonumber\\
&& +\frac{a_sN_t^4N_f}{4\xi^3}\left[\left. \frac{\partial m_f}{\partial a_s}\right|_{\xi}
\langle\overline{\psi}\psi\rangle+\left. \frac{\partial \nu_t}{\partial a_s}\right|_{\xi}\langle\overline{\psi}
\slash{D}_t\psi\rangle\right.\nn\\
&&+\left.\left.\frac{\partial \xi_o^{-1}}{\partial a_s}\right|_{\xi}\langle\overline{\psi}\slash{D}_s\psi\rangle\right].
\end{eqnarray}  

In order to be able to calculate numerically Eq.~(\ref{eq:23}) and Eq.~(\ref{eq:24}) we need to measure all the lattice observables in the above equations and to determine the
values of the Karsch coefficients 
$\left.\frac{\partial \beta}{\partial \xi}\right|_{a_s}$,  
$\left.\frac{\partial\xi_o}{\partial \xi}\right|_{a_s}$,
$\left. \frac{\partial m_f}{\partial \xi}\right|_{a_s}$,
$\left. \frac{\partial \nu_t}{\partial \xi}\right|_{a_s}$, 
$\left. \frac{\partial \beta}{\partial a_s}\right|_{\xi}$, 
$\left. \frac{\partial\xi_o}{\partial a_s}\right|_{\xi}$, 
$\left. \frac{\partial m_f}{\partial a_s}\right|_{\xi}$ and
$\left. \frac{\partial \nu_t}{\partial a_s}\right|_{\xi}$.

\fi


\section{Simulations and Scale Settings}
\label{sec:scales}

\ifnum\theScales=1
%
%
\ifnum\theOutline=1
\noindent \framebox{Begin \ outline \hspace{5.0in} }
\begin{enumerate}
\item Run table
\item Scale setting technique, step-size test, tuning of the velocity of light
\item Renormalized anisotropy from Wilson loops and $\rho$ masses 
\end{enumerate}
\noindent \framebox{End \ outline \hspace{5.0in} }
\vspace{0.25in}
\fi
For the purpose of our simulations we implement the $R$ algorithm \cite{R-alg} with step-size $\Delta t=0.005$ 
and stopping condition $10^{-6}$. Figure~\ref{fig:dt} shows that our choice for the step-size allows us 
to measure physical quantities with an error due to finite step-size smaller than 2\%, and that we are 
running in the stable regime of the $R$ algorithm.
For the spectrum measurements we use box sources of size 2 and local sinks
for all runs.

Our simulations examine the phase transition and the thermodynamic properties of QCD
 for volumes, temperatures, spatial lattice scales and quark masses similar to the already used
in the $N_t=4$, 2-flavor thermodynamic studies on isotropic lattices \cite{isotr_res}.
 We adjusted the bare parameters in the action so that the resulting physical 
anisotropy $\xi\approx4$, while $a_s\approx0.3$ fm and $m_\pi/m_\rho\approx 0.3$, which allowed the critical $N_t\approx 16$.

Another important step in tuning the bare parameters is choosing $\nu_t$ such that 
the relativistic properties of the anisotropic staggered action are restored. The velocity 
of light $c_{ts}$ is defined through the meson dispersion relation:
\begin{eqnarray}
E_{t, {\rm phys}}^{2}(P_{s, {\rm phys}})&=& \frac{E_{t, {\rm lat}}^{2}(0)}{a_t^2}+ c_{ts}^2P^{2}_{s, {\rm lat}}\frac{1}{a_s^2}, \nonumber
\end{eqnarray}
 where $E$ and $P$ are the energy and the momentum of the meson, subscripts ``lat'' and ``phys'' refer to a quantity in 
lattice or physical units, and $s$ and $t$ whether it is measured in the spatial or temporal
 direction. We tune $\nu_t$, so that the velocity of light $c_{ts}(P_{s, {\rm phys}}) \approx 1$. 
The velocity of light is calculated for the $\pi$ propagating in the temporal direction 
with a non-zero momentum for three valence values of this parameter, $\nu^{\rm val}_t=  0.8$, 1.0 and 1.2
(the dynamical value is $\nu^{\rm dyn}_t=1.0$). 
Figure~\ref{fig:c} demonstrates that the choice of $\nu^{\rm val}_t= 1.0$ gives a velocity of light closest to 1.0 for 
the set of bare parameters  $\beta=5.3$, $\xi_o=3.0$ and $m_f=0.008$. 

In Table~\ref{tab:nu_t} we compare masses measured in the temporal and spatial directions for
various combinations of $\nu^{\rm dyn}_t$ and $\nu^{\rm val}_t$. This shows that
a 20\% change in $\nu^{\rm dyn}_t$ has only a small effect compared to a
similar change in $\nu^{\rm val}_t$.
For the  masses we measure the essential contribution comes from $\nu^{\rm val}_t$.

Table~\ref{tab:runtable0}, Table~\ref{tab:runtableT1} and  Table~\ref{tab:runtableT2} list all the zero temperature scale setting runs 
and the finite temperature runs that we have done. The anisotropy $\xi$ for each of the zero temperature runs is calculated from the ratio of 
the $\rho$ masses in the spatial and temporal directions. For runs 1--5 and 7, Table~\ref{tab:runtable0},
 $\xi$ is determined from the matching 
of the static potentials as well (Figure~\ref{fig:pot_match} illustrates the matching technique \cite{wil_action}).
 The comparison between the two methods can be done examining Figure~\ref{fig:scatter},
which shows that they give reasonably close results. 

The quality of our data for all runs from Table~\ref{tab:runtable0} can be judged 
by studying the effective mass plots for $\pi$ and $\rho$. On Figure~\ref{fig:eff_pi_T2} through Figure~\ref{fig:eff_r_Z2}
we show some typical effective mass plots.
The effective mass at a given 
time or space slice is calculated from the values of the correlators at 2 neighboring points for $\pi$ and 4 for $\rho$, 
in order to determine all the parameters in the respective one and two cosh fitting forms. Ideally after some minimal time or 
space slice the effective mass plots should exhibit a plateau. For some effective mass plots determined from the spatial correlators
the quality of the plateau is not high, especially for the runs with very coarse lattice spacing, which means that 
for those runs there are larger errors on the meson masses determined from the fits to all data points.

All the data which we used in the Karsch coefficients and the EOS determination is given in Table~\ref{tab:all}.

\fi


\section{Phase Transition}
\label{sec:pahesetr}

\ifnum\thePhaseTr=1
%
%
\ifnum\theOutline=1
\noindent \framebox{Begin \ outline \hspace{5.0in} }
\begin{enumerate}
\item Result
\item Comparison with isotropic case
\end{enumerate}
\noindent \framebox{End \ outline \hspace{5.0in} }
\vspace{0.25in}
\fi
The finite temperature runs from Table~\ref{tab:runtableT1} and Table~\ref{tab:runtableT2} correspond 
to two sweeps through the phase transition for two different anisotropies $\xi=4.0(1)$ and $\xi=4.8(3)$, 
with corresponding $a_s$ of $0.34(1)$ fm, and $0.354(9)$ fm and $(m_\pi/m_\rho)^{temporal}$ 
of $0.33(1)$ and $0.325(8)$, respectively.

For each sweep through the transition region the temperature is changed only by changing $N_t$, while all
other parameters are kept constant. We want to stress the fact that there are no scale changes between the
different finite temperature runs from a group with a given anisotropy and that the scales change minimally 
between the two groups of runs belonging to the two different anisotropies. 

Figure~\ref{fig:pbp} shows the temperature dependence of $\langle \overline{\psi}\psi\rangle$ in the critical region.
From the data we can estimate $T_c\approx 150 -160$ MeV.
An interesting observations is that the shape of the transition is 
comparable in sharpness with the phase transition obtained in previous 
isotropic calculation \cite{iso_data,iso_data1}
also shown on the same figure. The isotropic data is from a dynamical 
staggered calculation with two fermion flavors 
on $16^3\times 4$ volume and $m_f=0.025$. The scale used to calculate the 
temperature in the isotropic case is from \cite{isotr_res}. 
The differences between our anisotropic result and the isotropic one we attribute 
to the scaling violations in the latter which we have avoided in our fixed parameter scheme.

\fi


\section{Karsch Coefficients}
\label{sec:karsch}

\ifnum\theKarsch=1
%
%
\ifnum\theOutline=1
\noindent \framebox{Begin \ outline \hspace{5.0in} }
\begin{enumerate}
\item Calculation technique: linear and quadratic fits in parameter space
\item Results
\end{enumerate}
\noindent \framebox{End \ outline \hspace{5.0in} }
\vspace{0.25in}
\fi

To determine the EOS we need to know the Karsch coefficients which are involved in the analytic 
expressions Eq.~(\ref{eq:23}) and Eq.~(\ref{eq:24}). 
The values of these derivatives can be calculated using the physical quantities that we measure 
for each zero-temperature run : $a_s$, $\xi$, $R_t=(m_\pi^2/m_\rho^2)^{\rm temporal}$ 
and $R_{st}=(m_\pi^2/m_\rho^2)^{\rm spatial}/(m_\pi^2/m_\rho^2)^{\rm temporal}$.
We consider the bare parameters $\xi_o$, $\beta$, $m_f$ and $\nu_t$ to be functions of the above 
physical quantities, which allows those functions to be expanded in Taylor series around the physical quantities of a selected 
zero-temperature run as follows:
\bea
\label{eq:fit1}
\Delta \xi_o(\xi, a_s, R_t, R_{st},\{\bf{c_i}\})&=& c_1 \Delta \xi + c_2 \Delta a_s + c_3 \Delta R_t +  c_4\Delta R_{st} + \cdots \\
\label{eq:fit2}
\Delta \beta(\xi, a_s, R_t, R_{st},\{\bf{d_i}\})&=& d_1 \Delta \xi + d_2 \Delta a_s + d_3 \Delta R_t +  d_4\Delta R_{st} + \cdots \\
\label{eq:fit3}
\Delta m_f(\xi, a_s , R_t, R_{st},\{\bf{e_i}\})&=& e_1 \Delta \xi + e_2 \Delta a_s + e_3 \Delta R_t +  e_4\Delta R_{st} + \cdots \\
\label{eq:fit4}
\Delta \nu_t(\xi, a_s, R_t, R_{st},\{\bf{f_i}\})&=& f_1 \Delta \xi + f_2 \Delta a_s + f_3 \Delta R_t +  f_4\Delta R_{st} + \cdots, 
\eea 
where 
$\Delta \xi_o = \xi_o - \xi^\prime_o $,
$\Delta \beta = \beta-\beta^\prime$,
$\Delta m_f = m_f -m^\prime_f$,
$\Delta \nu_t = \nu_t - \nu^\prime_t $,
$\Delta \xi   = \xi - \xi^\prime$,
$\Delta a_s   = a_s - a_s^\prime$,
$\Delta R_t    = R_t- R^\prime_t$,
$\Delta R_{st}    = R_{st} - R^\prime_{st}$.
In the last definitions the primed quantities refer to the selected run around whose physical quantities the 
Taylor expansion is done. The derivatives $c_i$, $d_i$, $e_i$ and $f_i$, $i=(1,\dots,4$), are defined as:
\begin{small}
\bea
\label{eq:kmatrix}
\hspace*{-13mm}\left( 
\begin{tabular}{cccc}
$c_1$&$c_2$&$c_3$&$c_4$\\
$d_1$&$d_2$&$d_3$&$d_4$\\
$e_1$&$e_2$&$e_3$&$e_4$\\
$f_1$&$f_2$&$f_3$&$f_4$\\
\end{tabular}
\right)&=&
\left( 
\begin{tabular}{rrrr}
\vspace{0.1cm}
$\left.\frac{\partial\xi_o}{\partial \xi}\right|_{a_s, R_t, R_{st}}$&
$\left.\frac{\partial\xi_o}{\partial a_s}\right|_{\xi, R_t, R_{st}}$&
$\left.\frac{\partial\xi_o}{\partial R_t}\right|_{\xi, a_s, R_{st}}$&
$\left.\frac{\partial\xi_o}{\partial R_{st}}\right|_{\xi, a_s, R_t}$ \\
\vspace{0.1cm}
$\left.\frac{\partial\beta}{\partial \xi}\right|_{a_s, R_t, R_{st}}$&
$\left.\frac{\partial\beta}{\partial a_s}\right|_{\xi, R_t, R_{st}}$&
$\left.\frac{\partial\beta}{\partial R_t}\right|_{\xi, a_s, R_{st}}$&
$\left.\frac{\partial\beta}{\partial R_{st}}\right|_{\xi, a_s, R_t}$ \\
\vspace{0.1cm}
$\left.\frac{\partial  m_f}{\partial \xi}\right|_{a_s, R_t, R_{st}}$&
$\left. \frac{\partial m_f}{\partial a_s}\right|_{\xi, R_t, R_{st}}$&
$\left.\frac{\partial  m_f}{\partial R_t}\right|_{\xi, a_s, R_{st}}$&
$\left.\frac{\partial  m_f}{\partial R_{st}}\right|_{\xi, a_s, R_t}$ \\
\vspace{0.1cm}
$\left.\frac{\partial\nu_t}{\partial \xi}\right|_{a_s, R_t, R_{st}}$&
$\left. \frac{\partial\nu_t}{\partial a_s}\right|_{\xi, R_t, R_{st}}$&
$\left.\frac{\partial\nu_t}{\partial R_t}\right|_{\xi, a_s, R_{st}}$&
$\left.\frac{\partial\nu_t}{\partial R_{st}}\right|_{\xi, a_s, R_t}$ \\
\end{tabular}
\right).
\eea
\end{small}
The Karsch coefficients, which are involved in the EOS, are the first two columns of the matrix of derivatives above.
We assume that we can make linear fits to Eq.~(\ref{eq:fit1}) through Eq.~(\ref{eq:fit4}) for 
each zero-temperature run and minimize the $\chi^2_i$, $i=(1,\dots,4)$, 
for all of the zero-temperature runs at the same time.
The $\chi^2_i$'s for the four fits are:

\bea
\label{eq:chi1}
\chi^2_1(\{{\bf c_i}\}) &=& \sum_r [\Delta \xi_o^r -  \Delta \xi_o^r(\xi^r, a_s^r, R_t^r, R_{st}^r,\{{\bf c_i}\})]^2/\sigma_r^2(\Delta \xi_o^r)\\
\label{eq:chi2}
\chi^2_2(\{{\bf d_i}\}) &=& \sum_r [\Delta \beta^r -  \Delta \beta^r(\xi^r, a_s^r, R_t^r, R_{st}^r,\{{\bf d_i}\})]^2/\sigma_r^2(\Delta \beta^r)\\
\label{eq:chi3}
\chi^2_3(\{{\bf e_i}\}) &=& \sum_r [\Delta m_f^r -  \Delta m_f^r(\xi^r, a_s^r, R_t^r, R_{st}^r,\{{\bf e_i}\})]^2/\sigma_r^2(\Delta m_f^r)\\
\label{eq:chi4}
\chi^2_4(\{{\bf f_i}\}) &=& \sum_r [\Delta \nu_t^r -  \Delta \nu_t^r(\xi^r, a_s^r, R_t^r, R_{st}^r,\{{\bf f_i}\})]^2/\sigma_r^2(\Delta \nu_t^r).
\eea
In the above expressions the sums are over $r$, which labels each zero-temperature run and the bare parameters 
and physical quantities associated with it. This labeling is not the same as the numbering of the runs 
in Table~\ref{tab:runtable0}, where 
each run number refers to a specific set of dynamical bare parameters.   
Here the subscript $r$ labels a specific set of bare parameters which instead of $\nu^{\rm dyn}_t$
has the valence value of that parameter, since as we already showed, the $\nu^{\rm val}_t$ has the dominant
contribution to the measured physical quantities. Hence in all formulae in this section the notation $\nu_t$ 
stands for the valence value of that parameter. 

All expansions are taken around a given selected run, whose label is not shown. 
The minimization of $\chi^2_1$ for the first fit, Eq.~(\ref{eq:fit1}) for example, leads to a matrix equation of the form 
\begin{equation}
\label{eq:mateq}
AC=V,
\end{equation}
where
\bea
A&=&\left(
\begin{tabular}{cccc}
\vspace{0.2cm}
$\sum_r \frac{\Delta \xi^r \Delta \xi^r}{\sigma_r^2(\Delta \xi_o^r)}$ &  
$\sum_r \frac{\Delta \xi^r \Delta a_s^r}{\sigma_r^2(\Delta \xi_o^r)}$ &
$\sum_r \frac{\Delta \xi^r \Delta R_t^r}{\sigma_r^2(\Delta \xi_o^r)}$ & 
$\sum_r \frac{\Delta \xi^r \Delta R_{st}^r}{\sigma_r^2(\Delta \xi_o^r)}$ \\
\vspace{0.2cm}
$\sum_r \frac{\Delta a_s^r \Delta \xi^r}{\sigma_r^2(\Delta \xi_o^r)}$ &  
$\sum_r \frac{\Delta a_s^r \Delta a_s^r}{\sigma_r^2(\Delta \xi_o^r)}$ & 
$\sum_r \frac{\Delta a_s^r \Delta R_t^r}{\sigma_r^2(\Delta \xi_o^r)}$ & 
$\sum_r \frac{\Delta a_s^r \Delta R_{st}}{\sigma_r^2(\Delta \xi_o^r)}$  \\
\vspace{0.2cm}
$\sum_r \frac{\Delta R_t^r \Delta \xi^r}{\sigma_r^2(\Delta \xi_o^r)}$ &  
$\sum_r \frac{\Delta  R_t^r \Delta a_s^r}{\sigma_r^2(\Delta \xi_o^r)}$ &
$\sum_r \frac{\Delta  R_t^r \Delta R_t^r}{\sigma_r^2(\Delta \xi_o^r)}$ & 
$\sum_r \frac{\Delta  R_t^r \Delta R_{st}^r}{\sigma_r^2(\Delta \xi_o^r)}$  \\
\vspace{0.2cm}
$\sum_r \frac{\Delta R_{st}^r \Delta \xi^r}{\sigma_r^2(\Delta \xi_o^r)}$ &  
$\sum_r \frac{\Delta  R_{st}^r \Delta a_s^r}{\sigma_r^2(\Delta \xi_o^r)}$ &
$\sum_r \frac{\Delta  R_{st}^r \Delta R_t^r}{\sigma_r^2(\Delta \xi_o^r)}$ & 
$\sum_r \frac{\Delta  R_{st}^r \Delta R_{st}^r}{\sigma_r^2(\Delta \xi_o^r)}$
\normalsize
\end{tabular}
\right)\nn,
\eea
\bea
\hspace*{-7.3cm}
V&=&\left(
\begin{tabular}{c}
\vspace{0.2cm}
$\sum_r \frac{\Delta \xi_o^r \Delta \xi^r}{\sigma_r^2(\Delta\xi_o^r )}$ \\
\vspace{0.2cm} 	
$\sum_r \frac{\Delta \xi_o^r \Delta a_s^r}{\sigma_r^2(\Delta \xi_o^r)}$ \\
 \vspace{0.2cm}	
$\sum_r \frac{\Delta \xi_o^r \Delta R_t^r}{\sigma_r^2(\Delta \xi_o^r)}$ \\	
\vspace{0.2cm}
$\sum_r \frac{\Delta \xi_o^r \Delta R_{st}^r}{\sigma_r^2(\Delta \xi_o^r)}$\\ 
\end{tabular}
\right)\nn,
\eea
\bea
\hspace*{-9cm}
C&=&\left(
\begin{tabular}{c}
$c_1$\\
$c_2$\\
$c_3$\\
$c_4$\\
\end{tabular}
\right)\nn,
\normalsize
\eea
which we solve for $C$. In a similar way we find matrix equations for the rest of the derivatives from Eq.~(\ref{eq:fit2}), Eq.~(\ref{eq:fit3}) and 
Eq.~(\ref{eq:fit4}).

To apply statistical analysis on our data from the zero-temperature runs, we divide the data into a set of jackknife blocks.
The numerical procedure for the minimization of the $\chi^2_i$ functions can not be applied straightforwardly for the equations
Eq.~(\ref{eq:chi1}) through Eq.~(\ref{eq:chi4})
since the standard deviations of the bare parameters, $\sigma_r^2(\dots)$, are not known from the beginning. Instead we employ an iterative 
scheme which consists of the following steps:
\begin{itemize}
\item Start by guessing initial values for all $\sigma_r^2(\dots)$'s.
\begin{enumerate}
\item Determine the Karsch coefficients by solving the matrix equation: Eq.~(\ref{eq:mateq}), and the similar equations 
derived from minimizing Eq.~(\ref{eq:chi2}) through Eq.~(\ref{eq:chi4}).
\item Using the values of the Karsch coefficients from step 1 calculate numerically the linear 
part of the functions Eq.~(\ref{eq:fit1}) through Eq. (\ref{eq:fit4}) 
on each jackknife block of data and by statistically analyzing them find new values for all $\sigma_r^2(\dots)$'s. 
\end{enumerate}
\item Repeat steps 1 and 2 until the numerical result for the Karsch coefficients converges.
\end{itemize}

The success of this scheme depends on how well the functions Eq.~(\ref{eq:fit1}) through Eq.~(\ref{eq:fit4}) 
can be approximated by the linear part of the Taylor expansion, which is 
a measure of how ``close'' the physical quantities measured from each run are to the quantities of the selected run around which the expansion 
is made.
To trust the consistency of the iterative scheme we checked the two step procedure with variety of random initial guesses for
 the $\sigma_r^2(\dots)$'s, which reproduced the same final results.

The numerical results for the Karsch coefficients from expansion around
runs 7 and 8 from Table~\ref{tab:runtable0}, obtained via the the method described above, are summarized in Tables~\ref{tab:kar1} and ~\ref{tab:kar2}. 
The quoted errors are calculated using the jackknife method. Our results show larger errors on the Karsch coefficients which are derivatives 
with respect to $a_s$ than the errors on those coefficients that are derivatives with respect to $\xi$. A possible explanation of that 
difference could be that the specific parameter space that we explored in our zero-temperature runs does not allow a better resolution 
of some of the Karsch coefficients either because it is too limited (we need more runs and more 
statistics on each of them to improve the quality of mass fits) or because the ``points'' in that space
(the zero-temperature runs) are not distributed in a favorable way around the run around which we are making the expansion, or both.

\fi


\section{Equation of State}
\label{sec:eos}

\ifnum\theEOS=1
%
%
\ifnum\theOutline=1
\noindent \framebox{Begin \ outline \hspace{5.0in} }
\begin{enumerate}
\item EOS results
\item Comparison with ideal gas 
\end{enumerate}
\noindent \framebox{End \ outline \hspace{5.0in} }
\vspace{0.25in}
\fi

In the previous section we described the procedure which allows us to calculate the Karsch coefficients
needed to determine the QCD equation of state (Eq.~(\ref{eq:23}) and Eq.~(\ref{eq:24})).

 As stressed before in Section~IV   
we have two groups of finite temperature runs listed in Tables~\ref{tab:runtableT1} and \ref{tab:runtableT2}
, for each of which we are changing 
the temperature by only varying $N_t$ and keeping the underlying physics scales fixed.
Figures~\ref{fig:E} and \ref{fig:P} show the numerical results for the energy density and pressure 
for both groups of runs corresponding to anisotropies $\xi=4.0(1)$ and $4.8(3)$.
The data is normalized to the continuum Stefan-Boltzmann values of the EOS for an ideal relativistic 
gas for $SU(N_c)$ color with $N_f$ quark flavors, which are 
\bea
\frac{\varepsilon_{\rm SB}}{T^4}&=&\frac{\pi^2}{15}(N_c^2-1+\frac{7}{4}N_cN_f)\approx 12.1725\nn
\eea
and
\bea
\hspace{-5cm}\frac{p_{\rm SB}}{T^4}&=&\frac{\varepsilon_{\rm SB}}{3T^4}.\nn
\eea

The errors on the pressure are significantly larger than the errors on the energy due to the
large errors on those of the Karsch coefficients which are derivatives with respect to $a_s$
and are involved only in Eq.~(\ref{eq:24}) for the pressure.
 The comparison with the free lattice theory (squares) gives an
explanation of the prominent drop off of $\varepsilon$ and $p$ in the
high temperature sector -- simply a consequence of the 
lattice high momentum cut-off. The high momentum mode contribution to the EOS becomes dominant
 with the increase of the temperature, which means that at a coarse lattice spacing a high proportion of 
relevant modes are simply excised. 
Including improvements  
to the spatial parts of the staggered fermion action would be a natural step to
reduce those lattice artifacts for high temperatures. 

Our results for the EOS are comparable with the isotropic case \cite{milcNt6} in the temperature region
up to 0.3 GeV for which the cited reference has data. The errors on the energy density are comparable with 
the errors in that isotropic study, or smaller with enough statistics, on the other hand the errors on pressure 
are larger due to the reasons stated above.  
\section{Note on Flavor Symmetry Improvement}
In the continuum and chiral limits the spontaneous symmetry breaking of $SU(4)_A\otimes SU(4)_V$ in
the staggered action, yields 15 Goldstone pions. On the lattice the violation of the flavor symmetry
leaves us with the remnant $U(1)_A\otimes U(1)_V$ and only one true Goldstone pion. The local pions in the staggered
formulation, which fall into 7 irreducible representations, are not degenerate any more due to the $O(a^2)$
flavor symmetry breaking. However the introduction of anisotropy on the lattice makes the
lattice spacing in the temporal direction much smaller than the spatial one 
and hence we expect to see an improvement in the flavor symmetry.

We choose $\Delta_\pi=(m_{\pi_2} - m_\pi)/m_\rho$, where $\pi_2$ is the second local staggered pion,
as a quantitative measure of the flavor symmetry breaking 
in the spatial and temporal directions. The data in Table~\ref{tab:flav}
shows that in the temporal direction for all runs $\Delta_\pi$ is smaller than its value in the spatial direction, 
which means that we are seeing improvement of the flavor symmetry as $a_t$ becomes finer.
Especially for run 4, the $\pi$ and $\pi_2$ look virtually degenerate.

We expect that the anisotropy has a similar effect on the rest of the pions, 
although we have not investigated numerically
how the various mass splittings between them are affected by the decrease of $a_t$.

\fi


\section{Conclusions}
\label{sec:conclusions}

\ifnum\theConclusions=1
%
%

We have studied the thermodynamic properties of full QCD with 2-flavors of staggered fermions 
on anisotropic lattices. In our calculations we have employed a fixed parameter scheme in which 
we keep the bare parameters constant and change the temperature by varying only the number of the 
temporal slices $N_t$. This allowed us to study the phase transition for staggered fermions with
fixed physics scales. It appears to be comparably as sharp as the transition in the isotropic 
case.

We have calculated non-perturbatively the Karsch coefficients from series of zero-temperatures runs and applied them 
in the determination of the EOS. Those of the Karsch coefficients which are derivatives with respect to $a_s$
have significant errors which are most probably due to the limited set of data used in their determination. 
They respectively give rise to large uncertainties in the calculation of the pressure. 

The high temperature behavior of the quark-gluonic system was found 
to be strongly influenced by the underlying lattice cut-off, which gives a maximum 
temperature at which our anisotropic EOS should represent continuum physics. However the fixed parameter 
scheme combined with a spatially improved anisotropic staggered action might give a much better 
result in the high temperature region.

The anisotropic approach naturally reduces the finite 
lattice spacing errors associated with $a_t$ and accounts 
for an improvement of the flavor symmetry for particles propagating 
in the temporal direction.

It is interesting to mention that our results do not show
a pronounced negative pressure problem in the confined phase as it has been found
in previous EOS calculations using the derivative method with perturbatively
calculated Karsch coefficients. However, considering the generally large
statistical errors on the pressure in our calculation, we could not entirely
exclude the possibility of such a problem being unveiled at low temperatures in a
calculation with
reduced statistical errors.

\section*{Acknowledgments}

We want to thank Lingling Wu for her significant contribution to the software used in this project
and George Fleming for useful discussions on the behavior of the free staggered fermion gas.
We also want to thank Norman Christ for his insightful ideas and much appreciated advice.
This work was conducted on the QCDSP machines at Columbia University and the RIKEN-BNL Research
Center. The authors are supported by the US DOE.

\fi

\newpage



\ifnum\theTables=1

\begin{table}[ptb]
\caption{Meson masses for run 1 and 2, Table~\ref{tab:runtable0}, measured with different valence and dynamical $\nu_t$'s. The
comparison shows that the main contribution to the masses comes from $\nu^{\rm val}_t$.
Notations $S$ and $T$ stand for spatial and temporal directions of measurement.}

\begin{center}
\begin{tabular}{lllllllllll}

{\em run}&  $\nu_t^{\rm dyn}$&  $\nu_t^{\rm val}$& {\em $m_\pi$, T} &{\em $m_\pi$, S} & {\em $m_\rho$, T} &{\em $m_\rho$, S}\\
\hline
1 & 1.0& 1.0& 0.31309(28) &  0.57846(83) & 0.6853(63)& 1.218(29)\\

2 & 1.2& 1.0& 0.31023(42) &  0.57677(46) & 0.6897(54)& 1.257(15) \\

2& 1.2& 1.2& 0.26761(35) &  0.58299(49) & 0.6030(43) & 1.232(32)\\
\end{tabular}
\end{center}
\label{tab:nu_t}
\end{table}

\begin{table}[ptb]
\caption{Parameters of zero-temperature calculations. All runs except run 2 have $\nu^{\rm dyn}_t=1.0$.
  Run 2 has $\nu^{\rm dyn}_t=1.2$.} 

\begin{center}
\begin{tabular}{llrlll} 

{\em run}& {\em volume} & {\em traj.}& $\beta$ & $\xi_o$ & $m_f$ \\ \hline

1 & $16^3\times32$ &  5800   & 5.425   & 1.5    &      0.025 \\ 

2 & $16^2\times24\times32$ & 5100 &  5.425 &     1.5      &  0.025\\

3 & $16^2\times24\times64$ & 1300 &  5.695  &  2.5   &   0.025\\ 

4 & $16^2\times24\times64$& 1400 &  5.725 &   3.44  &   0.025 \\ 

5 & $16^2\times24\times64$ & 3400 & 5.6 &   3.75   & 0.025 \\ 

6 &  $16^2\times24\times64$ & 4300 & 5.286 & 3.427 & 0.00394\\

7 & $16^2\times24\times64$ & 3200  & 5.3 & 3.0 & 0.008 \\  

8 & $16^2\times24\times64$ & 3000  & 5.29 & 3.4 &  0.0065 \\
\end{tabular}
\end{center}
\label{tab:runtable0}
\end{table}


\begin{table}[ptb]
\caption{Parameters of finite temperature calculations with $\xi=4.0(1)$. All runs have $\nu^{\rm dyn}_t=1.0$.}
\begin{center}
\begin{tabular}{llrlll} 
{\em run}& {\em volume} & {\em traj.}& $\beta$ & $\xi_o$ & $m_f$ \\ \hline
1 & $16^3\times24$ & 8000  & 5.3 & 3.0 &  0.008 \\ 

2 & $16^3\times20$ & 9800  & 5.3 & 3.0 & 0.008 \\  

3 & $16^3\times16$ & 21600     &5.3 & 3.0 &0.008\\

4 & $16^3\times12$ & 9100  & 5.3 & 3.0 & 0.008 \\

5 & $16^3\times8$ & 5500  & 5.3 & 3.0 &  0.008 \\

6 & $16^3\times4$ & 25900  & 5.3 & 3.0 &  0.008\\
 \end{tabular}
\end{center}
\label{tab:runtableT1}

\end{table}

\begin{table}[ptb]
\caption{Parameters of finite temperature calculations with $\xi=4.8(3)$. All runs have $\nu^{\rm dyn}_t=1.0$.} 
\begin{center}
\begin{tabular}{llrlll} 
{\em run}& {\em volume} & {\em traj.}& $\beta$ & $\xi_o$ & $m_f$ \\ \hline

1 & $16^3\times24$ & 1400  & 5.29 & 3.4 &  0.0065 \\

2 & $16^3\times20$ & 2700  & 5.29 & 3.4 &  0.0065 \\

3 & $16^3\times16$ & 8900  & 5.29 & 3.4 &  0.0065 \\

4 & $16^3\times12$ & 6200  & 5.29 & 3.4 &  0.0065 \\

5&  $16^3\times8$ & 3600  & 5.29 & 3.4 & 0.0065\\
\end{tabular}
\end{center}
\label{tab:runtableT2}
\end{table}


\begin{table}[ptb]
\caption{All zero-temperature data used to determine the Karsch coefficients 
from fits to Eq.~(\ref{eq:fit1}) -- Eq.~(\ref{eq:fit4}). 
The run index in the first column corresponds to the run number of Table~\ref{tab:runtable0},
 which lists the run parameters. Notations $S$ and $T$ stand for spatial and temporal directions of measurement.
$R_t$ and $R_{st}$ are defined in Section~V.} 
\label{tab:all}
\begin{center}
\begin{footnotesize}
\hspace*{-6mm}
\begin{tabular}{lllllllll}

run  & $\nu_t^{\rm val}$ & $\xi$ & $m_\pi, T$ & $m_\pi, S$ & $m_\rho, T$ & $m_\rho, S$ & $R_t$ & $R_{st}$\\\hline

1&  1.0 & 1.778(46) & 0.31309(28) &  0.57846(83) & 0.6853(63) & 1.218(29) & 0.2087(39) & 1.080(56)\\

2&  0.8 & 1.495(21) & 0.37050(41) &  0.56794(39) & 0.8227(69) & 1.230(14) & 0.2028(33) & 1.051(29)\\

2&  1.0 & 1.822(24) & 0.31023(42) &  0.57677(46) & 0.6897(54)& 1.257(15) & 0.2023(32) & 1.041(28)\\

2&  1.2 & 2.043(56) & 0.26761(35) &  0.58299(49) & 0.6030(43) & 1.232(32) & 0.1970(28) & 1.137(62)\\

3&  0.8 & 2.767(41) & 0.2451(10) &  0.6710(50) & 0.3144(34)& 0.870(11) & 0.608(12) & 0.979(27)\\

3&  1.0 & 3.637(80) & 0.1925(26) &  0.6774(48) & 0.2547(39) & 0.926(13) & 0.571(16) & 0.936(37)\\

3&  1.2 & 3.273(76) & 0.2094(39) &  0.6732(47) & 0.2716(54) & 0.889(13) & 0.594(12) & 0.966(31)\\

4&  0.8 & 3.722(41) & 0.2048(11) &  0.7643(24) & 0.2481(28) & 0.9235(58) & 0.681(16) & 1.005(23)\\

4&  1.0 & 4.295(54) & 0.17811(84) &  0.7699(24) & 0.2210(23) & 0.9493(62) & 0.649(14) & 1.013(27)\\

4&  1.2 & 4.888(70) & 0.15955(89) &  0.7786(23) & 0.1980(22) & 0.9680(65) & 0.649(14) & 0.997(28)\\

5&  0.8 & 3.990(34) & 0.20466(65) &  0.8456(27) & 0.2706(18) & 1.0796(74) & 0.5721(80) & 1.072(18)\\

5&  1.0 & 4.658(35) & 0.17661(76) &  0.8540(20) & 0.2382(13) & 1.1096(88) & 0.5497(80) & 1.078(16)\\

5&  1.2 & 5.260(46) & 0.15637(69) &  0.8624(22) & 0.21374(99) & 1.1242(90) & 0.5352(74) & 1.099(21)\\

6&  1.0 & 4.60(21) & 0.07413(25) &  0.36410(77) & 0.2811(39) & 1.294(51) & 0.0695(20) & 1.14(10)\\

7&  0.8 & 3.57(30) & 0.13273(42) &  0.47488(42) & 0.3777(51) & 1.35(11) & 0.1235(33) & 1.00(17)\\

7&  1.0 & 4.03(14) & 0.11280(51) &  0.47682(38) & 0.3274(39) & 1.320(49) & 0.1187(29) & 1.099(77)\\

7&  1.2 & 4.903(75) & 0.09800(44) &  0.47950(36) & 0.2857(34) & 1.401(15) & 0.1177(28) & 0.996(30)\\

8 & 1.0 & 4.80(28) & 0.09517(39) &  0.46121(58) & 0.2821(56) & 1.354(57) & 0.1138(45) & 1.02(10)\\
\end{tabular}
\end{footnotesize}
\end{center}
\end{table} 


\begin{table}[ptb]
\caption{The second local staggered pion masses $m_{\pi_2}$ and $\Delta_\pi=(m_{\pi_2}-m_\pi)/m_\rho$
from the zero-temperature runs with parameters given in Table~\ref{tab:runtable0}. 
The run index in the first column corresponds to the run number of Table~\ref{tab:runtable0}. 
Notations $S$ and $T$ stand for spatial and temporal directions of measurement.
For large values of the anisotropy $\xi$, $\Delta_\pi$ in the temporal direction is significantly smaller 
than the corresponding value in the spatial direction, for some runs even consistent with zero.}  

\begin{center}
\begin{small}
\begin{tabular}{lllllll}

run  & $\nu_t^{\rm val}$ & $\xi$ & $m_{\pi_2}$, $T$ & $m_{\pi_2}$, $S$ & $\Delta_\pi$, $T$ & $\Delta_\pi$, $S$\\
\hline
1&  1.0 & 1.778(46) &0.4605(37) &  1.109(25) &  0.2151(45) & 0.436(24)\\

2&  0.8 & 1.495(21) &0.5803(75) &  1.243(92) &  0.2551(89) & 0.549(79)\\

2&  1.0 & 1.822(24) & 0.4729(54) &  1.17(12) &  0.2359(75) & 0.475(95)\\

2&  1.2 & 2.043(56) & 0.3996(31) &  1.119(32) &  0.2188(54) & 0.435(27)\\

3&  0.8 & 2.767(41) & 0.2484(18) &  0.7462(75) &  0.0106(56) & 0.0864(87)\\

3&  1.0 & 3.637(80) & 0.1948(18) &  0.779(11) &  0.0091(65) & 0.110(10)\\

3&  1.2 & 3.273(76) & 0.2115(40) &  0.7592(80) &  0.008(11) & 0.0967(89)\\

4&  0.8 & 3.722(41) & 0.2071(11) &  0.8379(71) &  0.0094(42) & 0.0797(77)\\

4&  1.0 & 4.295(54) & 0.17931(97) &  0.8418(66) &  0.0054(47) & 0.0757(70)\\

4&  1.2 & 4.888(70) & 0.1598(10) &  0.8594(69) &  0.0011(59) & 0.0834(71)\\

5&  0.8 & 3.990(34) & 0.20853(83) &  0.9962(78) &  0.0143(29) & 0.1395(68)\\

5&  1.0 & 4.658(35) & 0.17954(77) &  1.0125(79) &  0.0123(31) & 0.1428(70)\\

5&  1.2 & 5.260(46) & 0.15877(75) &  1.0380(84) &  0.0112(33) & 0.1562(74)\\

6&  1.0 & 4.60(21) & 0.0956(14) &  0.58(11) &  0.0762(53) & 0.168(85)\\

7&  0.8 & 3.57(30) & 0.1693(14) &  1.30(36) &  0.0969(42) & 0.61(26)\\

7&  1.0 & 4.03(14) & 0.1410(11) &  1.24(16) &  0.0861(42) & 0.58(12)\\

7&  1.2 & 4.903(75) & 0.1213(13) &  1.181(79) &  0.0816(54) & 0.501(56)\\

8 & 1.0 & 4.80(28) & 0.1142(14) &  1.54(21) &  0.0674(54) & 0.80(16)\\
\end{tabular}
\end{small}
\end{center}
\label{tab:flav}
\end{table}


\begin{table}[ptb]
\caption{Karsch coefficients from fitting the data to the liner part of the Taylor expansion around run 7, 
Table~\ref{tab:runtable0}, with $\xi=4.0(1)$. The order of the coefficients is the same as in the matrix in Eq.~(\ref{eq:kmatrix}). Each row is obtained 
from the fitting procedure independently from the other rows and the $\chi^2$'s per degree of freedom for each fit is respectively
1.7, 1.0, 1.8 and 0.7.}
\begin{center}
\begin{tabular}{llll}
0.61(6) & 9.6(6.2)&  2.0(1.1)&  0.3(2.0)\\
-0.017(7)& -1.5(1.1) & 0.5(2)&  -0.03(28)\\
-0.0062(4)& 0.18(5)&  0.068(4) & -0.003(15)\\
0.04(5)& -5.6(4.8) & -0.7(8)&  1.1(1.4)\\
\end{tabular}
\end{center}
\label{tab:kar1}
\end{table}


\begin{table}[ptb]
\caption{Karsch coefficients from fitting the data to the liner part of the Taylor expansion around run 8, 
Table~\ref{tab:runtable0}, with $\xi=4.8(3)$. The order of the coefficients is the same 
as in the matrix in Eq.~(\ref{eq:kmatrix}). Each row is obtained 
from the fitting procedure independently from the other rows and the $\chi^2$'s per degree of freedom for each fit is
1.4, 0.7, 2.3 and 0.9.}
\begin{center}
\begin{tabular}{llll}
0.59(6)& 9.2(6.2)&  2.2(1.0) & 0.7(1.8)\\
-0.015(6)& -1.0(4) & 0.58(8)&  0.02(8)\\
-0.0050(8)& 0.11(7) & 0.05(1) & 0.02(2)\\
0.06(3) &-5.0(4.1) & -0.8(7)&  0.8(1.2)\\
\end{tabular}
\end{center}
\label{tab:kar2}
\end{table}

\fi


\ifnum\theFigures=1
%
%
\clearpage
\begin{figure}
\epsfxsize=\hsize
\begin{center}
\epsfbox{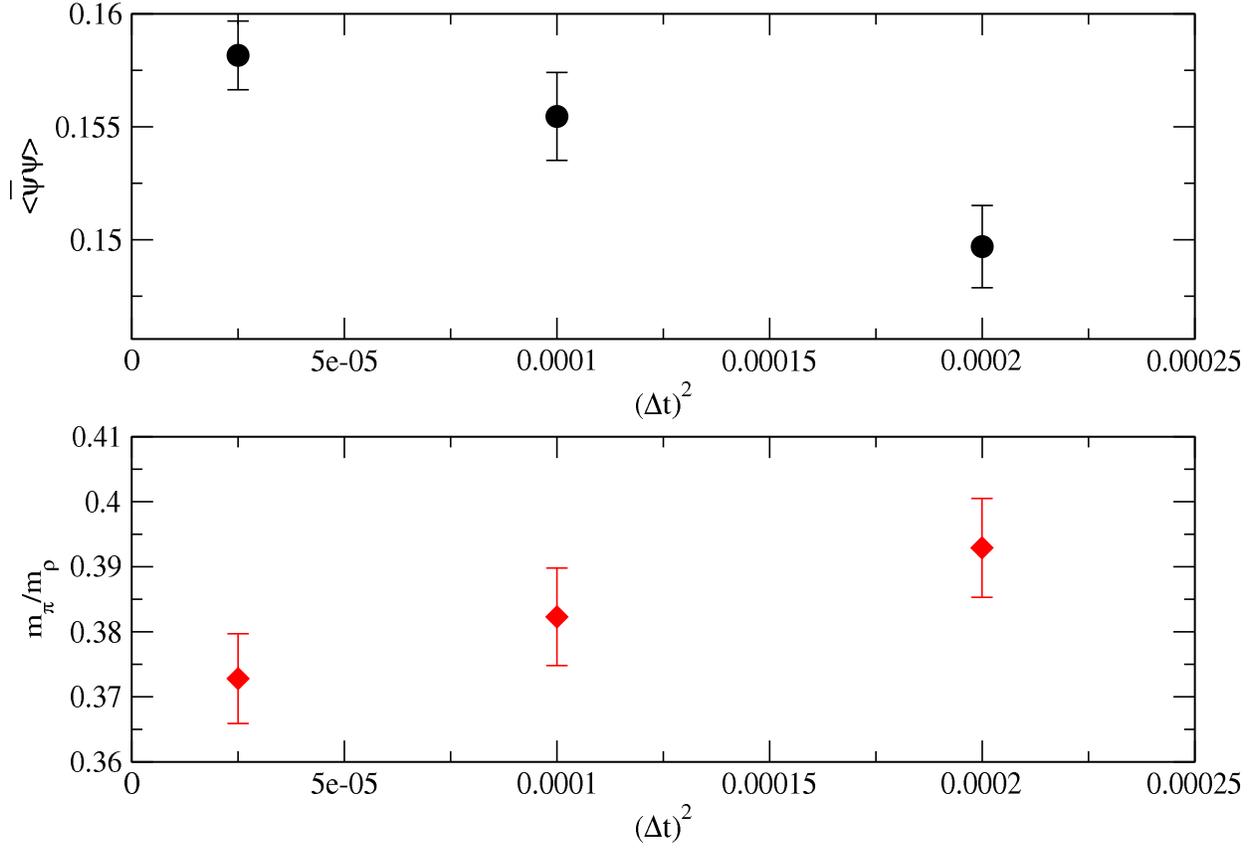}
\end{center}
\caption{Dependence of physical quantities on the step-size $\Delta t$ for 
volume $8^3\times 32$, $\beta=5.35$, $\xi_o=3.5$, $\nu^{\rm dyn}_t=\nu^{\rm val}_t=1.0$ and $m_f=0.006$.
Trajectories per point about 900. We simulate at $\Delta t =0.005$.}
\label{fig:dt}
\end{figure}

\clearpage
\begin{figure}
\epsfxsize=\hsize
\begin{center}
\epsfbox{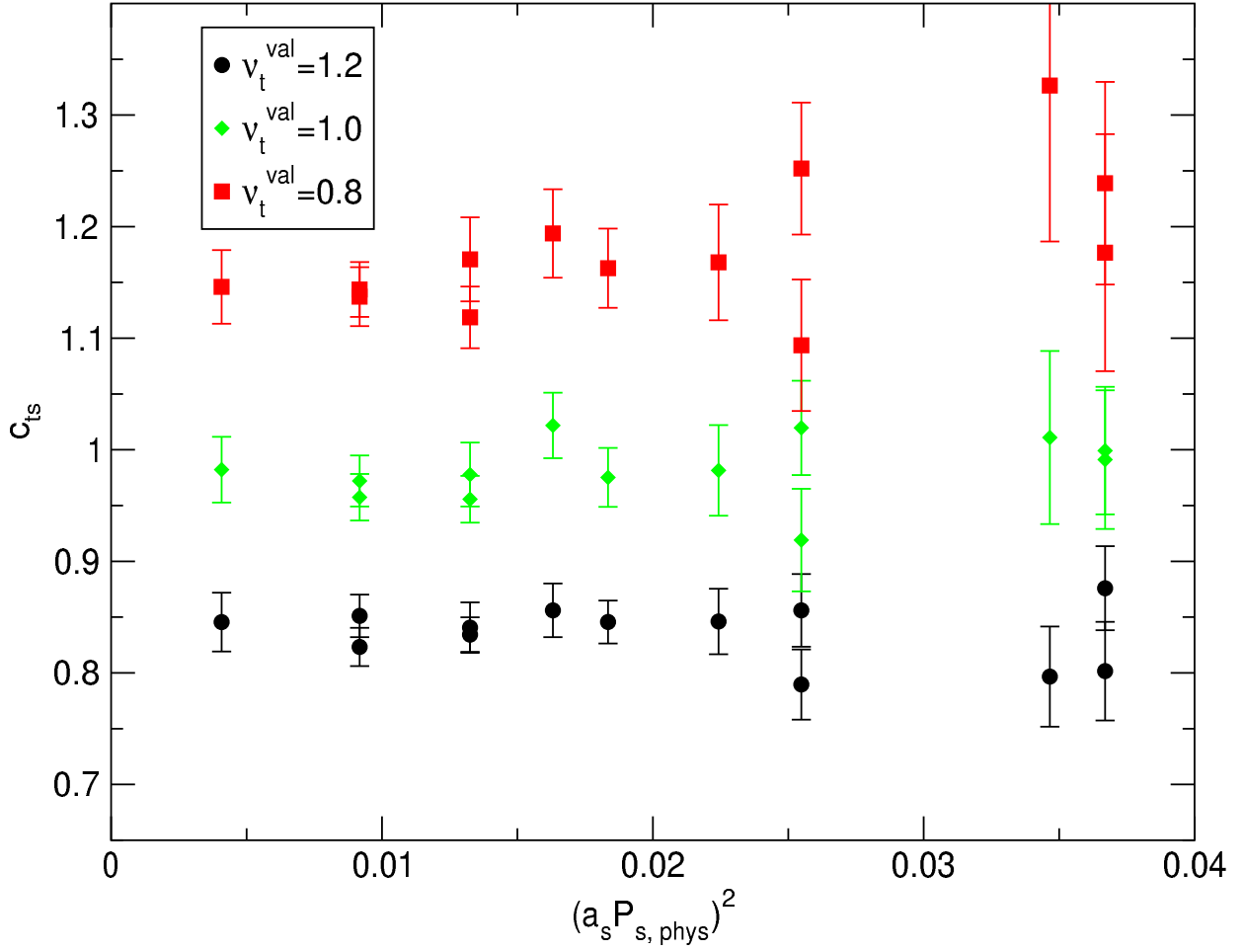}
\end{center}
\caption{Tuning of the velocity of light $c_{ts}$ using the dispersion relation in section~III 
for $\pi$. The run parameters are $\beta=5.3$, $\xi_o=3.0$, $m_f=0.008$, $\nu^{\rm dyn}_t= 1.0$; Measurements are done 
at  $\nu^{\rm val}_t= 0.8$, 1.0 and 1.2.
Square of the spatial momentum in lattice units $a_sP_{s, phys}$ is plotted on the horizontal axis.
The choice of $\nu^{\rm val}_t=1.0$ gives velocity of light closest to unity for that set of parameters.}
\label{fig:c}
\end{figure}

\clearpage
\begin{figure}
\epsfxsize=\hsize
\begin{center}
\epsfbox{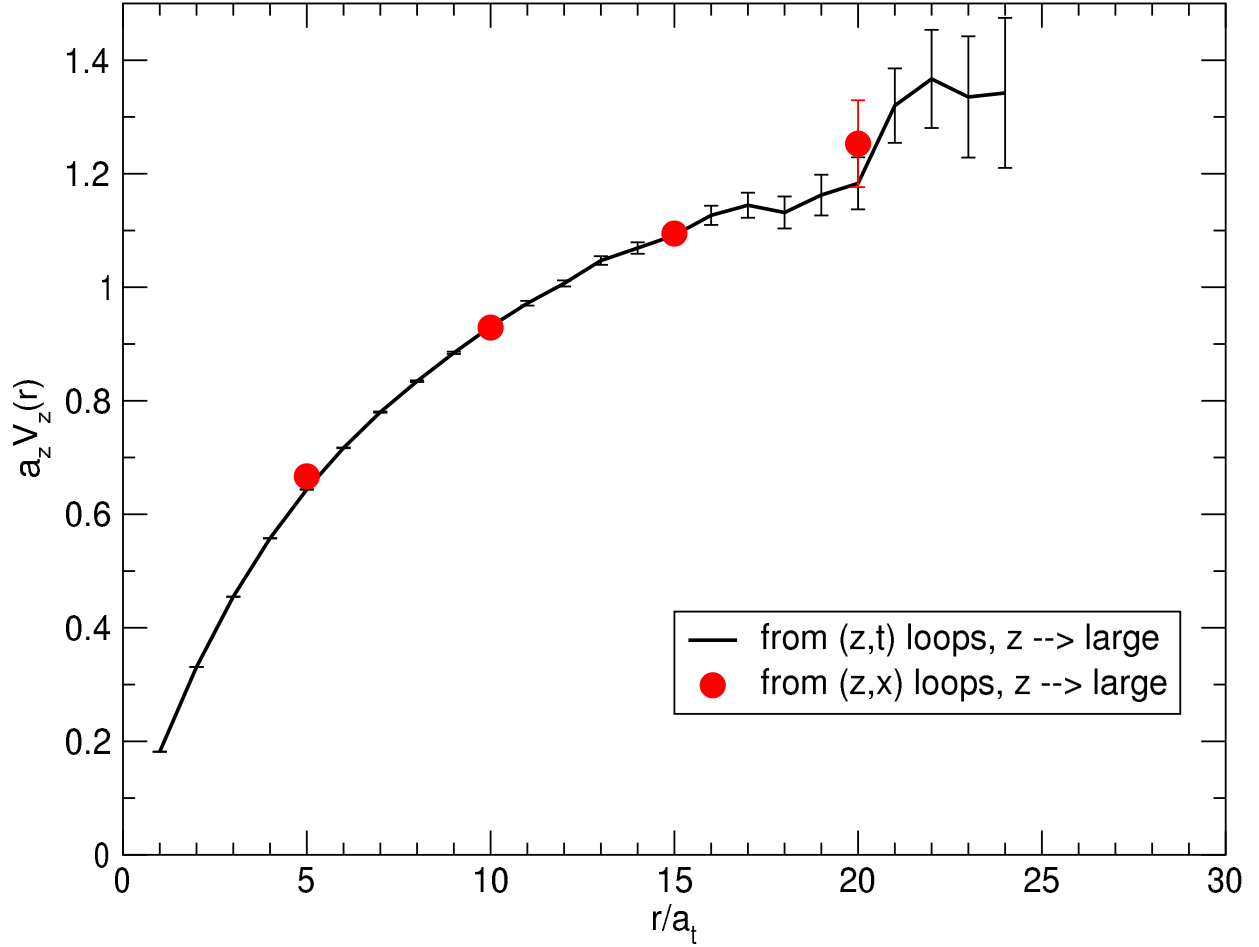}
\end{center}
\caption{Potential matching technique applied to static potentials measured from Wilson loops
in the $(z,x)$ and the $(z,t)$ planes for run 5, Table~\ref{tab:runtable0}. On this plot they are
shown after they are made to match by dividing the abscissa for the potential measured from the
$(z,x)$ loops by the anisotropy $\xi \approx 5$, so that $V_z(\xi a_t n)=V_z(a_s n)$.
}
\label{fig:pot_match}
\end{figure}

\clearpage
\begin{figure}
\epsfxsize=\hsize
\begin{center}
\epsfbox{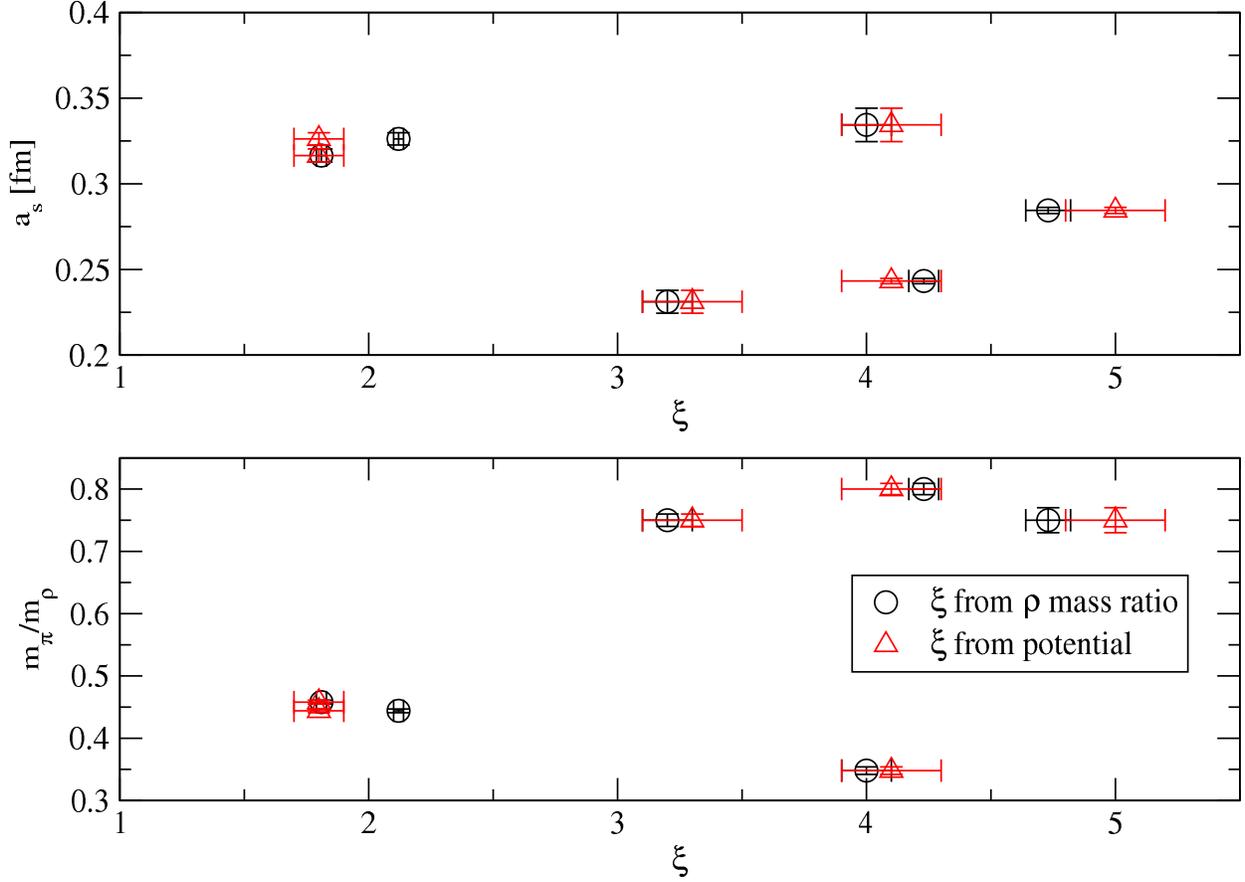}
\end{center}
\caption{Scatter plots for zero-temperature runs 1--5 and 7, Table~\ref{tab:runtable0}. 
The renormalized anisotropy $\xi$ is calculated both 
from the $\rho$ mass ratio in the spatial and temporal directions and from static potential matching.
The two methods give 
the same anisotropy but the errors for the potential matching results are larger.}
\label{fig:scatter}
\end{figure}



\clearpage
\begin{figure}
\epsfxsize=\hsize
\begin{center}
\epsfbox{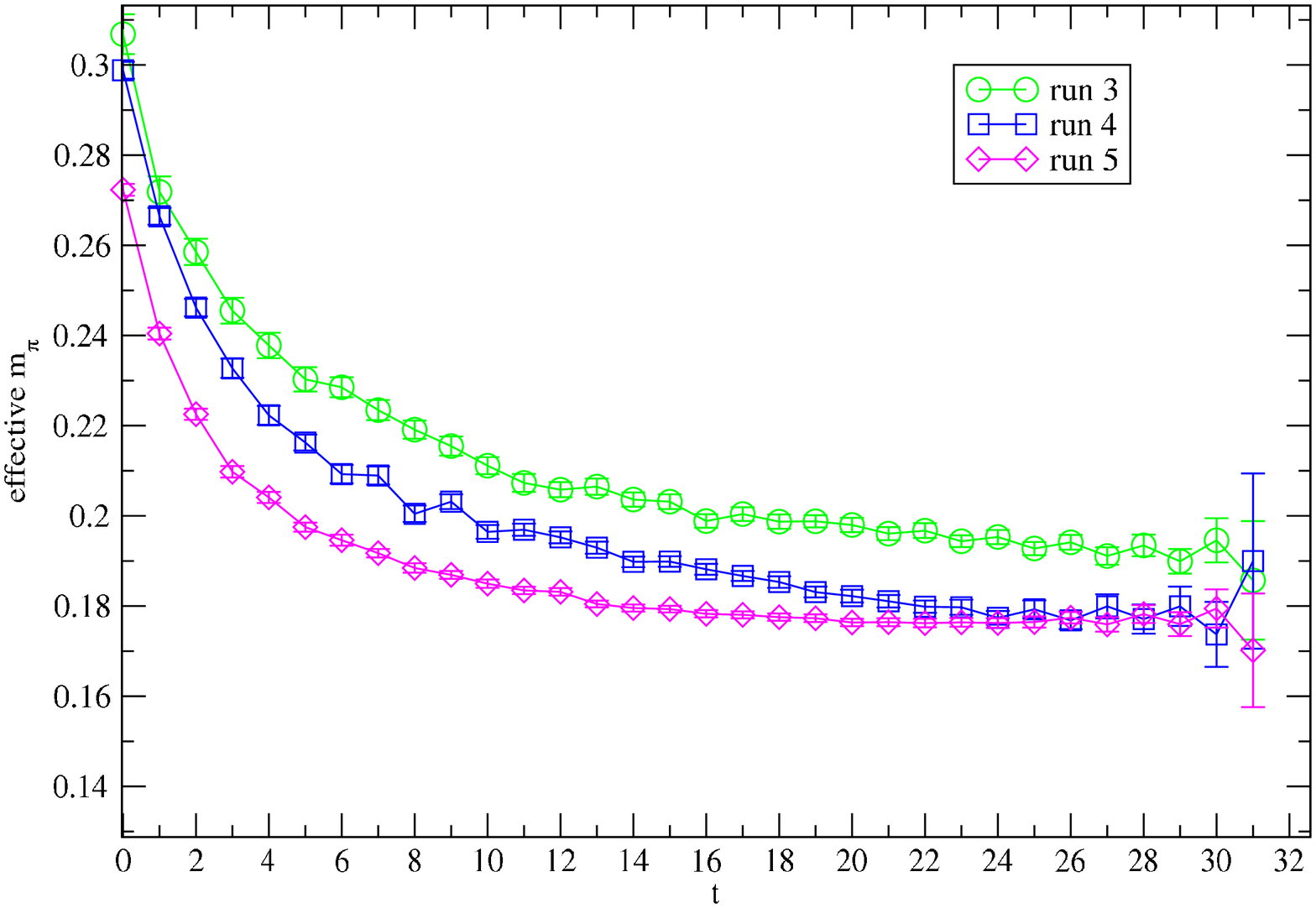}
\end{center}
\caption{Effective mass plot for $\pi$ propagating in temporal direction. Run parameters are 
as in Table~\ref{tab:runtable0} and data is measured at the dynamical value of $\nu_t$.}
\label{fig:eff_pi_T2}
\end{figure}

\clearpage
\begin{figure}
\epsfxsize=\hsize
\begin{center}
\epsfbox{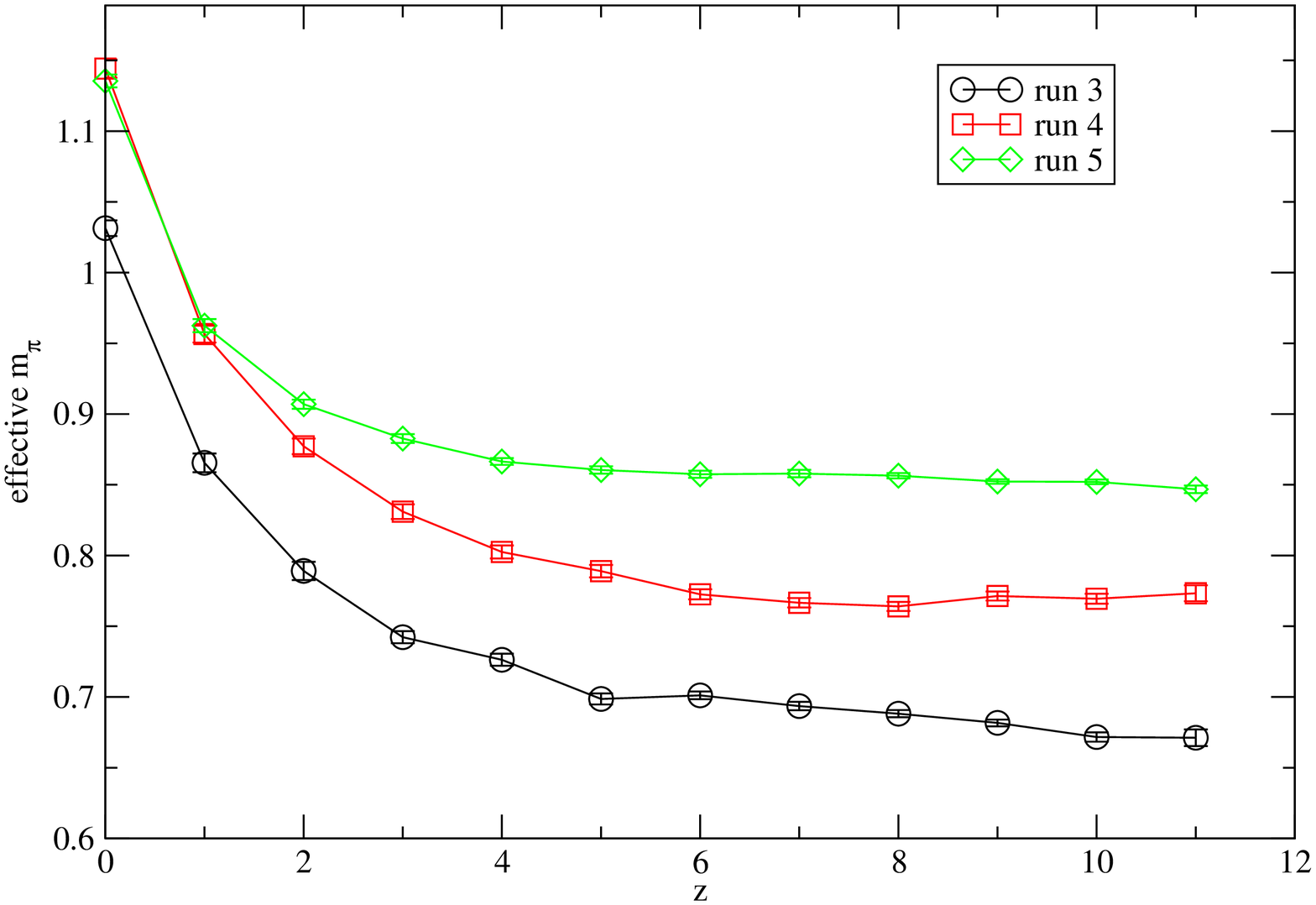}
\end{center}
\caption{Effective mass plot for $\pi$ propagating in spatial direction. Run parameters are 
as in Table~\ref{tab:runtable0} and data is measured at the dynamical value of $\nu_t$.}
\label{fig:eff_pi_Z2}
\end{figure}

\clearpage
\begin{figure}
\epsfxsize=\hsize
\begin{center}
\epsfbox{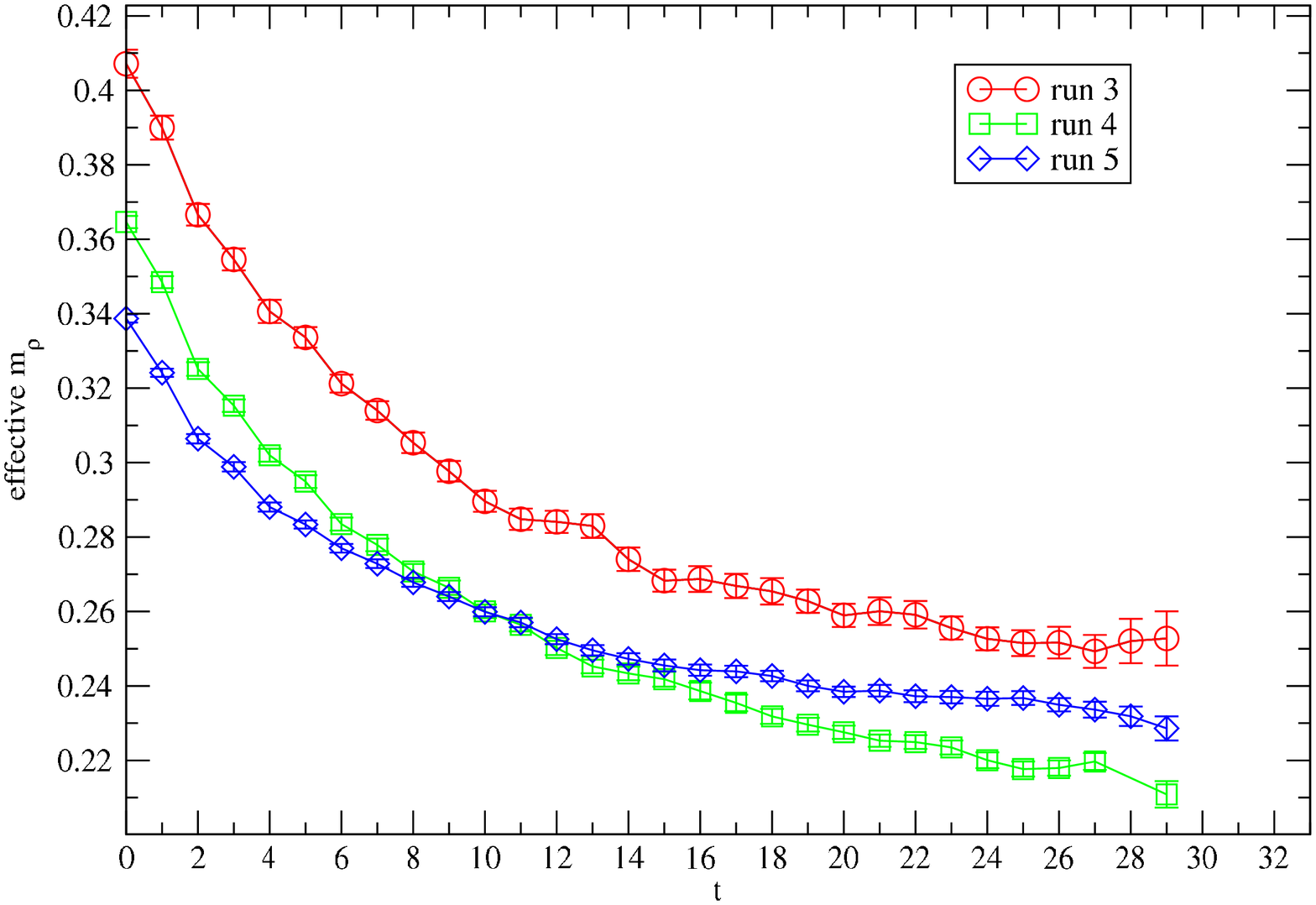}
\end{center}
\caption{Effective mass plot for $\rho$ propagating in temporal direction. Run parameters are 
as in Table~\ref{tab:runtable0} and data is measured at the dynamical value of $\nu_t$.}
\label{fig:eff_r_T2}
\end{figure}

\clearpage
\begin{figure}
\epsfxsize=\hsize
\begin{center}
\epsfbox{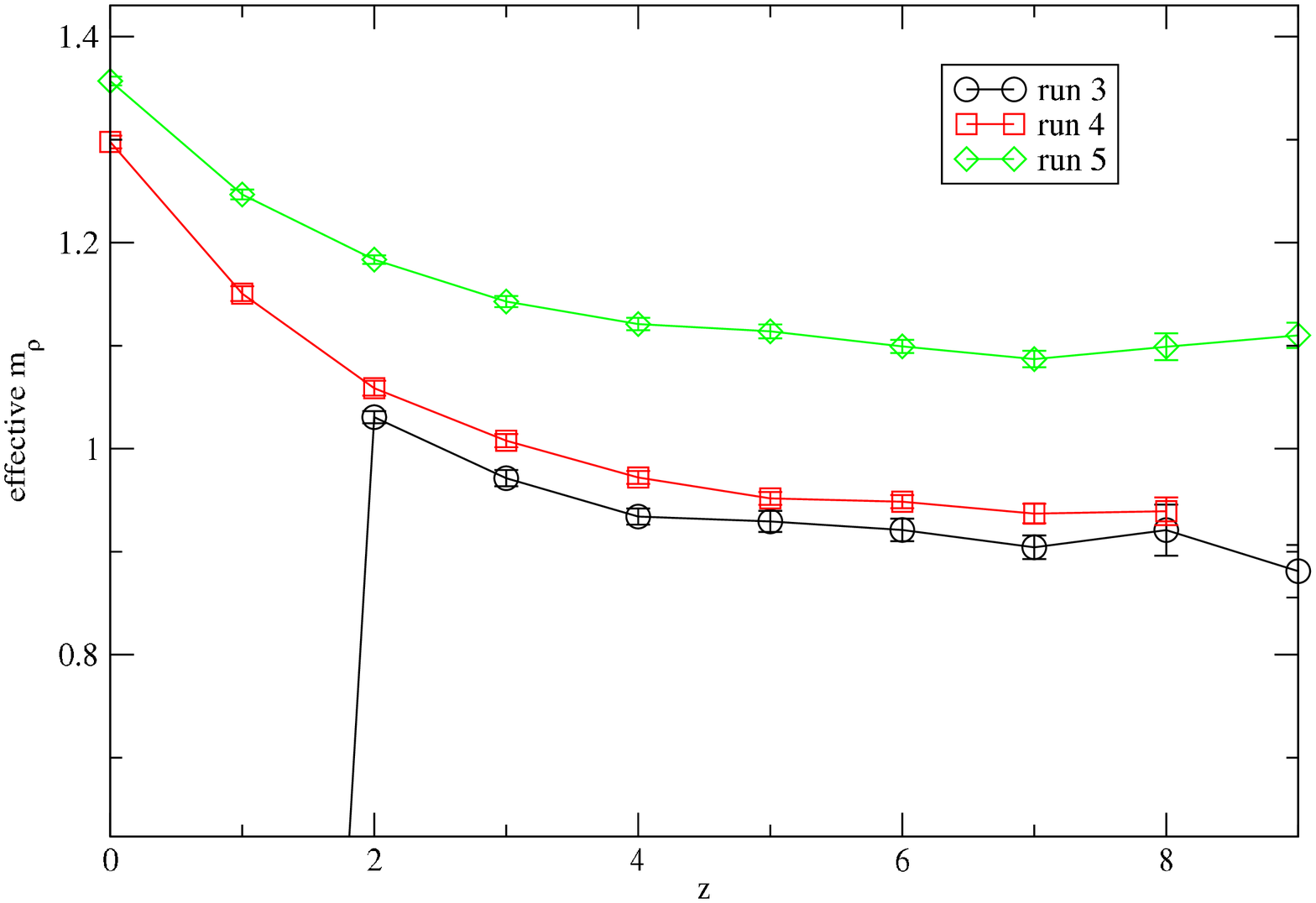}
\end{center}
\caption{Effective mass plot for $\rho$ propagating in spatial direction. Run parameters are 
as in Table~\ref{tab:runtable0} and data is measured at the dynamical value of $\nu_t$.}
\label{fig:eff_r_Z2}
\end{figure}

\clearpage
\begin{figure}
\epsfxsize=\hsize
\begin{center}
\epsfbox{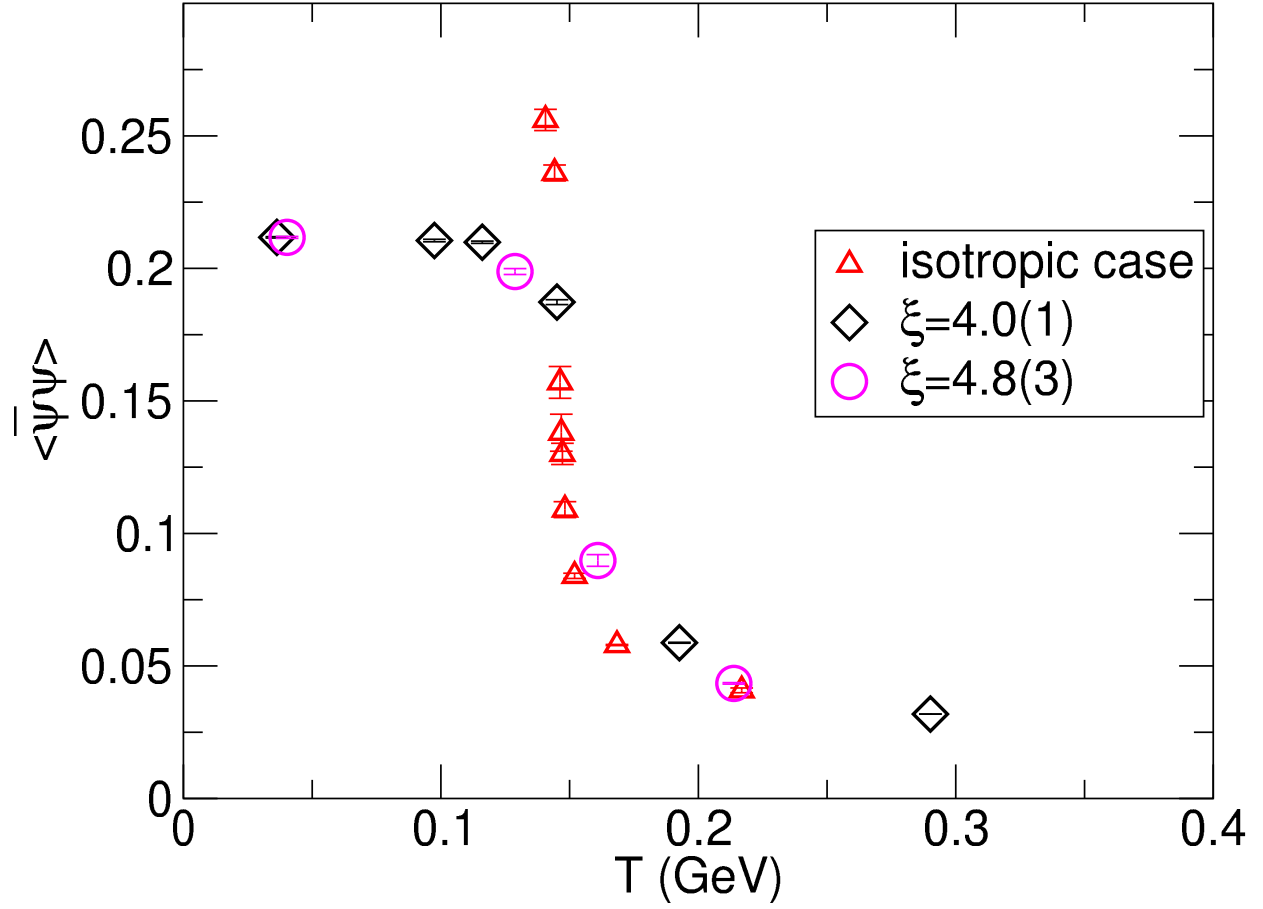}
\end{center}
\caption{
The temperature dependence of $\langle \overline{\psi}\psi\rangle$ in the region
of $T_c$. Points from anisotropic runs with a common symbol have the same anisotropy and physics scales. 
The isotropic data is shown for comparison.
From the critical region we estimate $T_c\approx 150$ -- 160 MeV.}
\label{fig:pbp}
\end{figure}

\clearpage
\begin{figure}
\epsfxsize=\hsize
\begin{center}
\epsfbox{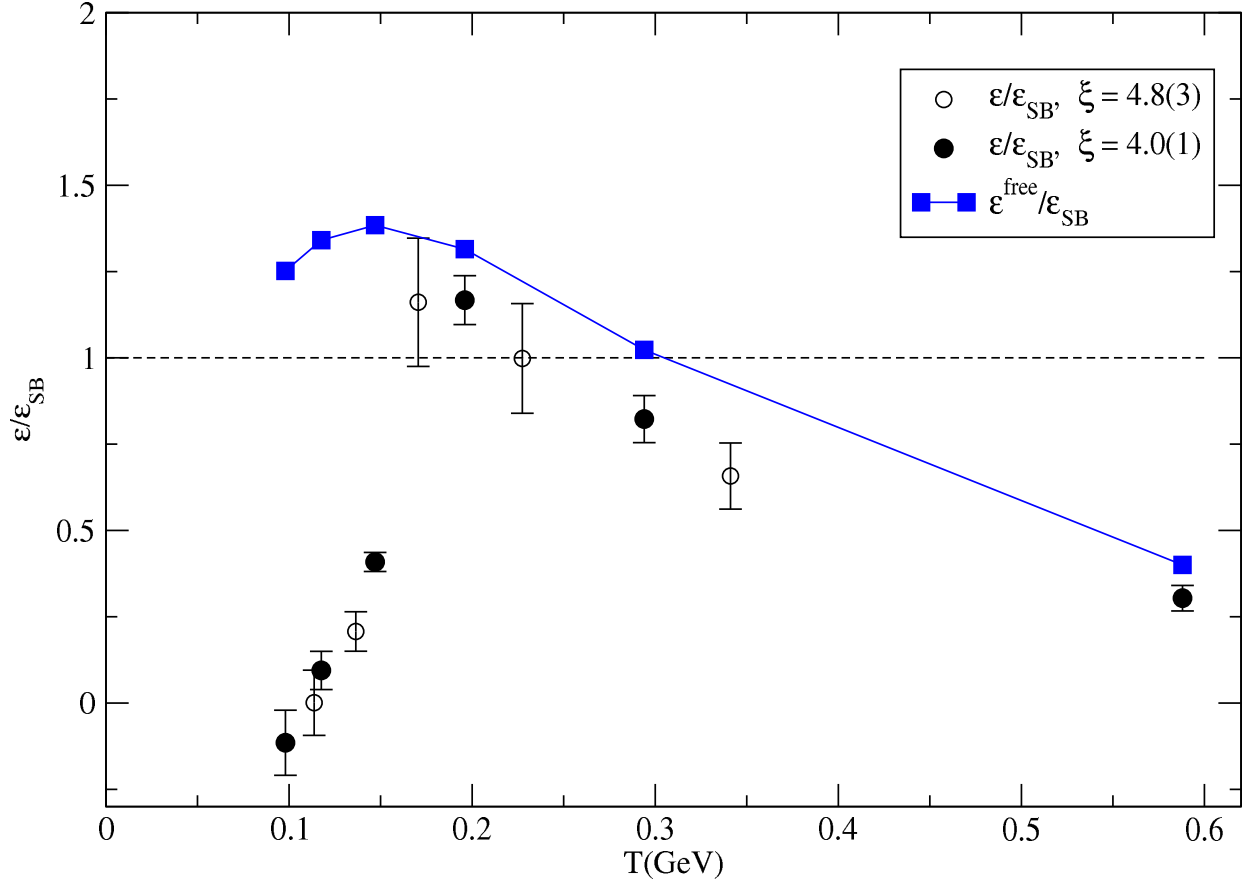}
\end{center}
\caption{Energy in units of the Stefan-Boltzmann limit and a comparison with
 the free lattice theory (squares). The Stefan-Boltzmann law for a relativistic ideal gas for $SU(N_c)$ color with $N_f$
 quark flavors in the continuum is 
$\frac{\varepsilon_{\rm SB}}{T^4}=\frac{\pi^2}{15}(N_c^2-1+\frac{7}{4}N_cN_f)\approx 12.1725$ for $N_c=3$ and $N_f=2$.
Points with a common symbol share the same anisotropy and the same physics scales at all temperatures.
}
\label{fig:E}
\end{figure}

\clearpage
\begin{figure}
\epsfxsize=\hsize
\begin{center}
\epsfbox{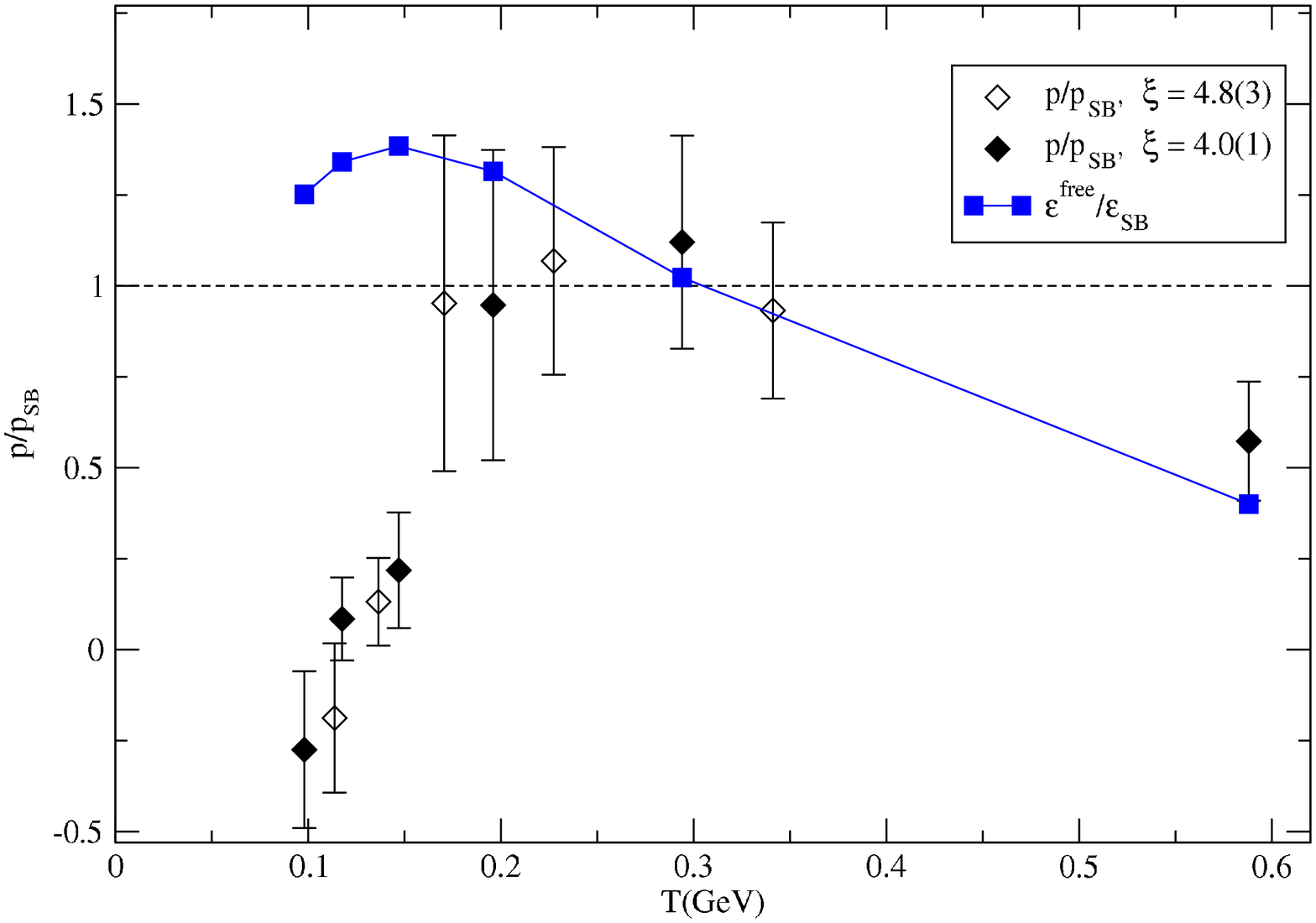}
\end{center}
\caption{Pressure in units of the Stefan-Boltzmann limit and a comparison with
 the free lattice theory (squares). The Stefan-Boltzmann law for a relativistic ideal gas for $SU(3)$ color with $N_f=2$
 quark flavors in the continuum gives $\frac{p_{\rm SB}}{T^4}=\frac{\varepsilon_{\rm SB}}{3T^4}$. Points with a common symbol 
share the same anisotropy and the same physics scales at all temperatures.}
\label{fig:P}
\end{figure}

\fi


\begin{thebibliography}{99}
\bibitem{k_coeff} F. Karsch, {\em Nucl. Phys.} {\bf B205[FS5]}, 285 (1982).\vspace{.5cm}
%
\bibitem{diff_method} G. Burgers et al., {\em Nucl. Phys.} {\bf B304}, 587 (1988).\vspace{.5cm}
%
\bibitem{karsch_pert} R. C. Trinchero, {\em Nucl. Phys.} {\bf B227}, 61 (1983).\vspace{.5cm}
%
\bibitem{karsch_pert1} F. Karsch and I. O. Stamatescu, {\em Phys. Lett.} {\bf B227}, 153 (1989).\vspace{.5cm}
%
\bibitem{isotr_res} T. Blum et al., {\em Nucl. Phys. (Proc. Suppl.)} {\bf B34}, 320 (1994);
{\em Phys. Rev.} {\bf D51}, 5153 (1995).\vspace{.5cm}
%
\bibitem{int_method} S. Huang et al., {\em Phys. Rev.} {\bf D42}, 2864 (1990).\vspace{.5cm}
%
\bibitem{int_method1} J. Engels et al., {\em Phys. Lett.} {\bf B252}, 625 (1990).\vspace{.5cm}
%
\bibitem{anis_lat} QCD-TARO Collaboration: Ph. de Forcrand et al., {\em Phys. Rev.} {\bf D63:054501} (2001).\vspace{.5cm}
%
\bibitem{wil_action} T. Klassen, {\em Nucl. Phys. }{\bf B533}, 557 (1998).\vspace{.5cm}
%
\bibitem{stagg_action} L. Susskind, {\em Phys. Rev.} {\bf D16}, 3031 (1977). \vspace{.5cm}
%
\bibitem{R-alg} S. Gottlieb et al., {\em Phys. Rev.} {\bf D35}, 2531 (1987).\vspace{.5cm}
%
\bibitem{iso_data} A. Vaccarino, {\em Nucl. Phys. (Proc. Suppl.)} {\bf B20}, 263 (1991).\vspace{.5cm} 
%
\bibitem{iso_data1} "Numerical Studies of the QCD Finite Temperature Phase Transition with Two and Three 
Flavors of Light Quarks", Ph.D. thesis by A. Vaccarino, Columbia University (1991).
%
\bibitem{milcNt6} MILC Collaboration: C. W. Bernard et al., {\em Phys. Rev.} {\bf D55}, 6861 (1997).\vspace{.5cm}
%
\end{thebibliography}
\end{document}